\documentclass[aip,reprint,floatfix]{revtex4-2}

\usepackage{graphicx}
\usepackage[breaklinks=true,colorlinks=true,linkcolor=blue,urlcolor=blue,citecolor=blue]{hyperref}
\usepackage{upgreek}
\usepackage{siunitx}

\begin{document}

\title{Inflationary stimulated Raman scattering in shock-ignition plasmas}

\author{S. J. Spencer}
\email[]{s-j.spencer@warwick.ac.uk}
\homepage[]{sj-spencer.github.io}
\affiliation{Centre for Fusion, Space, and Astrophysics, University of Warwick, Coventry, CV4 7AL}

\author{A. G. Seaton}
\affiliation{Centre for Fusion, Space, and Astrophysics, University of Warwick, Coventry, CV4 7AL}
\affiliation{Los Alamos National Laboratory, Los Alamos, New Mexico 87545, USA}

\author{T. Goffrey}
\affiliation{Centre for Fusion, Space, and Astrophysics, University of Warwick, Coventry, CV4 7AL}

\author{T. D. Arber}
\affiliation{Centre for Fusion, Space, and Astrophysics, University of Warwick, Coventry, CV4 7AL}

\date{\today}

\begin{abstract}
In the shock-ignition inertial confinement fusion scheme, high-intensity lasers propagate through an
inhomogeneous coronal plasma, driving a shock designed to cause fuel ignition. During the high-intensity
ignitor laser pulse, in the long scale-length
coronal plasma, back-scattered stimulated Raman scattering (SRS) is likely to be in the kinetic regime.
In this work, we use one-dimensional particle-in-cell simulations to
show that there is a non-linear frequency shift caused by kinetic effects, resulting in the growth
of SRS in an inhomogeneous plasma
far exceeding the predictions of fluid theory, so-called inflationary SRS or iSRS. We find that iSRS occurs
over a wide range of density scale-lengths relevant to shock-ignition and other directly-driven inertial
confinement fusion schemes. The presence of iSRS in shock-ignition plasmas
has implications for the theoretical gains from shock-ignition inertial
confinement fusion. Here we quantify the intensity threshold for the onset of iSRS for shock-ignition relevant
parameters.
\end{abstract}

\maketitle

\section{Introduction}

 Shock-ignition (SI) \cite{Betti_2008,Perkins2009} is a direct-drive \cite{Craxton2015} inertial confinement fusion
 (ICF) scheme where the fuel assembly and ignition stages of the implosion are decoupled using a characteristic laser
 pulse shape.
 In the first stage a low-intensity ($10^{14}-10^{15}$ \si{W/\centi\metre^2}) pulse is used to ablate the outer
 layer of the target, creating a long density scale-length coronal plasma and compressing the fuel.
 In the second stage, high-intensity ($10^{15}-10^{16}$ \si{W/\centi\metre^2}) beams are used to launch a
 strong shock into the target, which in turn leads to a non-isobaric pressure profile.
 In a successful experiment, when the pressure peaks at the centre, ignition occurs.
 In the ignition stage of the pulse, the average laser intensity on target is above the threshold for
 laser-plasma instabilities (LPI) \cite{Theobald2012}, such as stimulated
 Raman scattering (SRS)\cite{Liu1974}, stimulated Brillouin scattering (SBS)\cite{Liu1974}, and
 two-plasmon decay (TPD)\cite{Liu1976}.

 One laser-plasma instability of major concern\cite{Rosenberg2018} to shock-ignition  is stimulated Raman scattering, a three-wave parametric instability that transfers energy from the laser to an electron plasma wave (EPW)
 and a scattered light wave \cite{Liu1974}.
 SRS is only possible in plasma regions where the following matching conditions can be satisfied:
 $\omega_0 = \omega_\mathrm{EPW} + \omega_\mathrm{s}$, $\mathbf{k}_0 = \mathbf{k}_\mathrm{EPW} + \mathbf{k}_\mathrm{s}$,
 where the subscripts refer to the incident laser, electron plasma wave, and scattered light wave, respectively.
 The production of scattered light by SRS is deleterious to the SI scheme as it diverts laser energy away from
 the target and reduces laser-illumination uniformity.
 The effect of SRS-generated electron plasma waves on shock-ignition is less well understood, as it depends on
 the specific waves which are amplified. As the SRS EPWs Landau damp, they transfer energy to electrons with velocities
 $v \sim v_\mathrm{ph} = \omega/k > v_\mathrm{th}$.
 Of these suprathermal electrons, those with energies less than $ 100\mathrm{keV}$ are predicted to stop behind the
 converging shock, and could act to augment the ignition shock \cite{Ribeyre_2009}.
 Electrons with energies greater than $100\mathrm{keV}$ are likely to deposit their energy ahead of the shock and
 pre-heat the target, making it more difficult to compress \cite{Batani_2014}.
For this reason, it is important to understand the energy distribution of suprathermal electrons produced by SRS in shock-ignition scenarios.

Previous simulation studies of stimulated Raman scattering in shock-ignition have identified dominant SRS growth at
$n_\mathrm{cr}/4$ and $n_\mathrm{cr}/16$,
where it grows as an absolute instability\cite{Klimo2010,Klimo2011}. SRS has also been observed at densities such that
it grows as a weakly kinetic convective instability; characterised by the condition that\cite{Cristoforetti2017}:
$ 0.15 < k_\mathrm{EPW}\lambda_\mathrm{D} < 0.25$. Furthermore, some studies have shown the presence of SRS in the strongly kinetic limit
($k_\mathrm{EPW}\lambda_\mathrm{D} > 0.25$), where it occurs by kinetic inflation \cite{Riconda2011,Batani_2014}, as explained below.

SRS which occurs via kinetic inflation (from here on referred to as inflationary SRS, or iSRS) has been studied
extensively in the low-density homogeneous plasmas relevant to indirect-drive ICF on the NIF
\cite{Vu2002,Yin2006,Vu2007,Strozzi2007,Yin2008,Yin2012,Ellis2012}.
In attempting to explain experimental measurements of large SRS-reflectivities at high values of $k_\mathrm{EPW}\lambda_\mathrm{D}$
\cite{Fernandez2000,Montgomery2002}, it was suggested that some mechanism caused a reduction in the Landau damping
rate by four to five times, compared to the damping for a Maxwellian plasma \cite{Montgomery2002}.
Anomalously large SRS-reflectivities were recreated in simulations \cite{Vu2001,Vu2002}, and were explained by
reference to O'Neil's 1965 model of reduced EPW  damping caused by electron-trapping\cite{ONeil1965}.
In homogeneous plasmas, iSRS occurs when an SRS EPW grows to a point where it can trap electrons for one
complete bounce period or longer, without them becoming de-trapped due to velocity-space diffusion or
side-loss\cite{Vu2002}. This trapped electron population leads to modification of the distribution
function, in the form of a locally flattened region around the EPW phase velocity.
This translates to a modification of the dielectric properties of the plasma, resulting in a reduction in
the EPW's associated Landau damping rate \cite{ONeil1965,Vu2002}, and increased SRS growth.

 Key results of previous studies of inflationary SRS in homogeneous plasmas include: a theory for the saturation
 of iSRS in terms of EPW bowing and the trapped-particle modulation instability \cite{Yin2008}; the derivation
 of an inflation threshold intensity in terms of competition between trapping in the EPW and diffusion in
 velocity space \cite{Vu2007}; and the description of iSRS in terms of a transition from convective to absolute
 growth \cite{Wang2018}. Inflationary SRS has also been identified as an important mechanism in simulations  with ensembles of laser speckles \cite{Yin2012,Winjum2019}.

 In the large-scale inhomogeneous plasmas associated with shock-ignition ($L_n =n_e/(dn_e/dx) \simeq 300-1000 \si{\micro\metre}$) the mechanism and effects of inflationary SRS have, so far, received little attention. The few papers which do refer to iSRS in SI inhomogeneous plasmas assume that the explanation of iSRS in a homogeneous plasma in terms of reduced Landau damping also applies to iSRS in an inhomogeneous plasma. However, iSRS in an inhomogeneous plasma actually happens by a different mechanism. In the case of a homogeneous plasma, reduced Landau damping due to electron-trapping in the EPW leads to an increase of the SRS growth rate which, if sufficiently large, can cause a transition from convective to absolute growth \cite{Wang2018}. For an inhomogeneous plasma, where the growth of SRS is always convective, the reduction in Landau damping associated with trapping in the EPW has no net effect on the convective gain. While the local SRS growth rate may depend on the EPW damping rate in an inhomogeneous plasma, the region of SRS convective growth is also extended, leading to a net Rosenbluth gain\cite{Rosenbluth1972} which is independent of Landau damping \cite{Williams1991,Liu1994}. We therefore look to another non-linear effect caused by the trapped electrons in the SRS EPW, the non-linear frequency shift\cite{Morales1972}. In an inhomogeneous plasma, the frequency shift resulting from electron-trapping can compensate for the wave-number mismatch on propagating up the density gradient, thereby allowing
 growth over a larger region - an auto-resonance \cite{Chapman2010,Chapman2012}. Chapman {\it et al.} (2012) \cite{Chapman2012},  proposed this theory for iSRS in an inhomogeneous $(L_n \lesssim 100
 \si{\micro\metre})$ plasma close to the hohlraum wall in indirect-drive ICF. They demonstrated the auto-resonant
 interaction\cite{Chapman2010} between the non-linear frequency shift associated with electron-trapping in EPW
 and the wave-number mismatch caused by plasma inhomogeneity; which allows larger SRS gain\cite{Chapman2012}.

 Inflationary SRS has been suggested as the cause of SRS from low densities
 in simulations of LPI in shock ignition \cite{Klimo2014}. Sub-scale shock-ignition experiments have detected
 SRS scattered light from densities $0.09 - 0.16 n_\mathrm{cr}$, where the inflationary mechanism should be important \cite{Cristoforetti2017}. Recent full-scale ($L_n > 500
 \si{\micro\metre}$, $T_e = 5\si{\kilo \electronvolt}$) directly-driven experiments have detected significant SRS-reflected light from densities $0.15 - 0.21 n_{\mathrm{cr}}$ \cite{Rosenberg2020}. Another full-scale ($L_n = 450
 \si{\micro\metre}$, $T_e = 4.5\si{\kilo \electronvolt}$) SI experiment measured SRS-reflected light from densities between $0.05 - 0.15 n_{\mathrm{cr}}$. Under the conditions of the experiment, $k_{\mathrm{EPW}}\lambda_D$ ranges from $0.3 - 0.6$ and the measured SRS is assumed to be inflationary in origin \cite{Baton2020}.
 For a single laser speckle in an inhomogeneous plasma with density
 scale-length $L_n\simeq 70 \si{\micro\metre}$, Riconda {\it et al.} (2011) \cite{Riconda2011} demonstrated
 that iSRS  was associated
 with electron-trapping in the EPW. By varying $a_0=eE_0/c m_e \omega_0$ from 0.03 to 0.06, i.e. an increase in
 laser intensity from $1.0\times10^{16} \si{W/\centi\metre^2}$ to $4\times10^{16} \si{W/\centi\metre^2}$, they showed
 a transition to iSRS.

 The primary aim of this study is to identify shock-ignition plasma parameters where iSRS may occur. This information will guide
 future studies examining the longer term consequences of iSRS growth, such as its saturation mechanisms or
 interaction with other instabilities, which will require more detailed and computationally expensive modelling.
 Here we show that inflationary SRS can occur in inhomogeneous plasmas with density scale-lengths
 $L_n\simeq 300-1000 \si{\micro\metre}$, such as expected for shock-ignition coronal plasmas, and at
 shock-ignition laser intensities and plasma temperatures.
 We demonstrate that iSRS in
 these simulations is characterised by electron-trapping, frequency shift of the EPW and the appearance of beam-acoustic modes (BAM). Through a set of parameter studies we estimate the transition threshold
 for iSRS at different density scale-lengths and densities. These simulations are all restricted to 1D
 to determine where iSRS can occur in isolation from other effects.
 We also comment on the possible risk to SI from inflationary SRS and approaches
 to maximise the benefit of the hot-electron production.

 The outline of this paper is as follows: Section \ref{sec:code&IC} describes the code used and the choice of initial conditions. In Section \ref{sec:signatures} we demonstrate how one may detect the presence of iSRS in PIC simulations of shock-ignition plasmas. Section \ref{sec:paramScan} describes how the iSRS behaviour changes with the density scale-length of the coronal plasma. We also characterise the hot-electron populations, and comment on their potential impact on the ignition shock. Section
 \ref{sec:conclusion} summarises our results.


\section{Code and initial conditions}\label{sec:code&IC}

Simulations are performed using the particle-in-cell (PIC) code EPOCH \cite{Arber2015}, which solves Maxwell's equations on a
fixed grid and self-consistently moves particles under the Lorentz force law.
The initial plasma conditions span a range of SI-relevant parameters. The simulation parameters are chosen to achieve our primary aim of identifying plasma parameters where iSRS may occur, which does not require large simulations of the entire LPI system.

The simulations all used a domain size of $L_x = 100\si{\micro\metre} $ and ran to $T_\mathrm{end} = 2\si{\pico\second}$
with 2048
particles per cell (PPC) for the electron species.
We treat the ions as a neutralising background population since we simulate only a two pico-second interval of SRS
development, during which ion dynamics will not become important \cite{Rousseaux2006}.
For the plasma parameters laid out above, electron-ion collisions occur on a characteristic timescale of approximately
$7 \si{\pico\second}$ at the highest density probed, $0.22n_\mathrm{cr}$. Since the inflationary Raman process we
are investigating takes place
on a sub-picosecond timescale, we do not include collisions in our simulations.
The plasma density profiles are given by the expression $n(x) = n_\mathrm{min}\mathrm{exp}(x/L_n)$ and can be seen in Table
\ref{tab:densities}.

We simulate a frequency-tripled Nd:glass laser with vacuum wavelength $\lambda_0 = 351\si{\nano\metre}$, polarised in the $y$-direction. The laser intensity was varied in 20 logarithmically evenly-spaced increments between $10^{14}$\si{W/\centi\metre^2} and $10^{16}$\si{W/\centi\metre^2}, with a half-Gaussian temporal profile followed by a flat top, and a rise-time of 50 laser periods.
We use absorbing boundaries for the fields and thermal for the particles; these replace any particle leaving the
simulation with an incoming particle with velocity consistent with a Maxwellian plasma based on the initial temperature of $4.5$\si{keV}.

\begin{table}[ht]
    \caption{\label{tab:densities}
        Summary of density profiles and $k_{EPW}\lambda_\mathrm{D}$ values in each simulation. $L_n=n_e/(dn_e/dx)$ evaluated at $n_\mathrm{mid}$. For all but the case centred at $0.2n_\mathrm{cr}$, $k_\mathrm{EPW}\lambda_\mathrm{D} > 0.28$ and we are in the strongly kinetic regime. The total range of $k_\mathrm{EPW}\lambda_\mathrm{D}$ probed is 0.21-0.41.
        }
    \begin{ruledtabular}
    \begin{tabular}{cccc}
    $L_n/\si{\micro \metre}$  & $n_\mathrm{mid}/n_\mathrm{cr}$ & $(n_\mathrm{min},n_\mathrm{max})/n_\mathrm{cr}$ &$(k\lambda_\mathrm{D_{min}},k\lambda_\mathrm{D_{max}})$\\
    \hline
    300& 0.15  & $(0.13,0.18)$ & $(0.28,0.37)$\\
    500 & 0.12 &$(0.11,0.13)$ & $(0.37,0.41)$\\
    500 & 0.15 & $(0.14,0.17)$& $(0.29,0.35)$ \\
    500 & 0.20 & $(0.18,0.22)$& $(0.21,0.27)$\\
    1000 & 0.15 & $(0.14,0.16)$ & $(0.31,0.32)$ \\
    \end{tabular}
    \end{ruledtabular}
\end{table}

EPOCH uses a pseudorandom number generator (PRNG) to generate the initial particle distribution.
Each simulation was repeated 10 times with a different PRNG seed, allowing us to determine the sensitivity of
SRS to plasma fluctuations.
This allowed us to calculate both the mean and standard deviation of the intensity of the light scattered through SRS.

\begin{figure}[ht]
 \centering
 \includegraphics[width=\columnwidth]{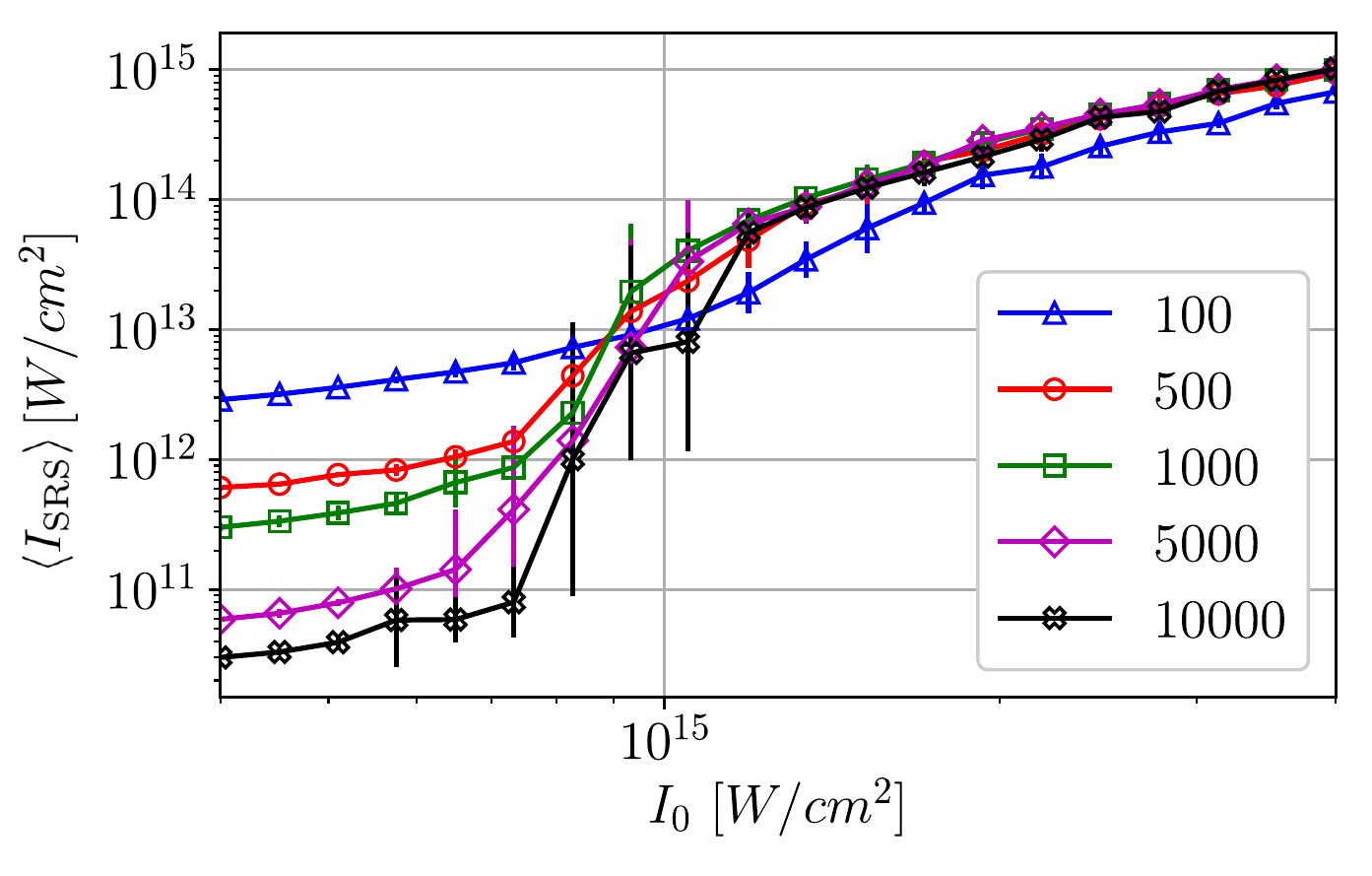}
 \caption{(Colour) Time-averaged intensity of SRS scattered light for a homogeneous simulation ($n_e=0.15n_\mathrm{cr}$, $T_e = 4.5$\si{\kilo \electronvolt}) with different numbers of particles
 per cell. Relative errors are given by one standard deviation of the SRS scattered light intensity as calculated from ten simulations.} \label{fig:convergence}
\end{figure}

SRS amplifies fluctuations in the plasma, and so is sensitive to number of particles per cell used in the simulation, as can be seen in Figure \ref{fig:convergence}.
In this figure, the intensity of SRS back-scattered light (denoted in this paper by $\langle I_{\mathrm{SRS}}\rangle$) is plotted against the incoming laser intensity $I_0$ ranging between
$0.4 - 4.0 \times 10^{15}$\si{W/\centi\metre^2} for a homogeneous plasma with $n_e=0.15n_\mathrm{cr}, T_e = 4.5$\si{\kilo \electronvolt}, for different numbers of
particles per cell.
At low incident intensity, $\langle I_{\mathrm{SRS}}\rangle$ is inversely proportional to the number of PPC used. This is as we
would expect for the case of simple convective amplification \cite{Rosenbluth1972} of the product of two quantities ($E_y,B_z$) which vary as background PIC noise, which is proportional to $1/\sqrt{\mathrm{PPC}}$.
The upper saturated level of  $\langle I_{\mathrm{SRS}}\rangle$ is robust to the number of PPC for $\mathrm{PPC} > 100$.
The transition between these two levels represents the change from standard convective amplification of SRS to enhanced
growth of SRS due to trapping (inflationary SRS), hence we call this the inflation threshold.
The existence of an inflation threshold is also robust to the number of particles per cell for $\mathrm{PPC} > 100$.

In the region containing the inflation threshold, the error associated with the intensity of SRS scattered light is largest. This suggests that inflationary SRS is very sensitive to the initial distribution of particles in the simulation domain, and that a statistical analysis of the mean and standard deviation of intensity across different random seeds will be important if we are to determine the iSRS threshold intensity accurately.


\section{Signatures of iSRS in inhomogeneous plasmas}\label{sec:signatures}

Three signatures of inflationary SRS observed in the literature for homogeneous plasmas are: a threshold intensity past which scattering of laser light is enhanced above the level predicted by fluid theory \cite{Vu2007}; electron-trapping in the SRS EPWs leading to local flattening of the distribution function at the EPW phase velocity \cite{Vu2002}; and the growth of down-shifted SRS EPWs and a continuum of beam-acoustic modes (BAMs) \cite{Yin2006}.
In what follows we show that all of these signatures are also present for iSRS in an inhomogeneous plasma, despite the instability arising through an auto-resonance rather than a transition from convective to absolute growth.

We consider first the existence of  a threshold intensity past which SRS growth is enhanced, by several
orders of magnitude, above the predictions of fluid theory; this has been seen in experiments \cite{Kline2006} and simulations \cite{Vu2002,Yin2006,Vu2007,Riconda2011}. In order to identify kinetic inflation of the SRS scattered light intensity in our PIC simulations,
we use a simple fluid model to calculate and compare the intensity of SRS scattered light in the absence of kinetic effects.

According to fluid theory, the growth of a parametrically unstable mode in an inhomogeneous plasma is limited by the loss of resonance between the waves as they propagate through the plasma and experience wave-number shift \cite{Rosenbluth1972}. We can formulate this inhomogeneous growth in terms of the Rosenbluth gain exponent\cite{Rosenbluth1972}
\begin{equation}\label{eqn:GRos}
    G_\mathrm{Ros} = 2\pi\gamma_0^2/|v_{g,1}v_{g,2}\kappa'|,
\end{equation}
where $\gamma_0$ is the growth rate of the equivalent mode in a homogeneous plasma, $v_{g,1}, v_{g,2}$ are the group speeds of the scattered EM wave/EPW and $\kappa'$ is the $x$-derivative of the wave-number mismatch $\kappa(x) = k_0(x) -k_\mathrm{s}(x) -k_\mathrm{EPW}(x)$. The maximum intensity reached by a parametrically unstable wave which has grown from noise at point $x$ is then given by the expression $I_\mathrm{noise}\mathrm{exp}(G_\mathrm{Ros}(x))$. In order to calculate the intensity of scattered light due to SRS, we substitute for $k_0,k_\mathrm{s},k_\mathrm{EPW}$ using the electromagnetic and Bohm-Gross dispersion relations in one dimension, to get:
\begin{equation}\label{eqn:kappaPrime}
    \frac{d\kappa}{dx}= -\frac{1}{2}\frac{q_e^2}{m_e\epsilon_0}
    \left(\frac{1}{c^2k_0}-\frac{1}{3v_\mathrm{th}^2k_\mathrm{EPW}}-\frac{1}{c^2k_\mathrm{s}}\right)\frac{dn_e}{dx}.
\end{equation}
Substituting this back into $G_\mathrm{Ros}$ with the growth rate for backward SRS in a homogeneous plasma\cite{kruer2003},
\begin{equation}\label{eqn:gamma0}
    \centering
    \gamma_0 = \frac{k_\mathrm{EPW}v_{os}}{4}\left[\frac{\omega_{\mathrm{pe}}^2}{\omega_\mathrm{EPW}(\omega_0-\omega_\mathrm{EPW})}\right]^{1/2},
\end{equation}
gives an appropriate Rosenbluth gain exponent for calculating convective amplification of
back-scattered SRS light in our simulations.

We make several simplifying assumptions that allow us to estimate the maximal scattered light intensity at a point in our simulation domain. Firstly, we neglect the dependence of the scattered light velocity on space, and consider it to be fixed at $c$. This means that we slightly over-estimate the amount of scattered light which is able to reach the point $x$ in time $t$. We also assume that the laser achieves its maximum intensity starting at $t=0$ rather than ramping up, as it does in the simulations.
We neglect collisional damping of the scattered EM waves and assume that the noise source $I_\mathrm{noise}$ is homogeneous in the domain. Finally, we assume that the amplification described by the Rosenbluth gain exponent occurs locally and instantaneously at the point of perfect
matching $(\kappa=0)$, rather than across the resonance region defined by $\ell \sim 1/\sqrt{\kappa'}$. For all the simulations presented in this paper $\ell < 6\si{\micro\metre}$.
 The scattered light intensity is then given by
 \begin{equation}\label{eqn:fluid_model}
     I(x) = \frac{1}{L_x}\int_x^{L_x}I_\mathrm{noise}\mathrm{exp}(G_\mathrm{Ros}(s)) ds.
 \end{equation}
The prefactor $1/L_x$ ensures that if $G_\mathrm{Ros}=0$, such that there is no growth, then the back-scattered signal remains
at the the noise intensity. The steady-state intensity of SRS scattered light measured at the laser-entry boundary is then given by $\langle I_{\mathrm{SRS}} \rangle = I(0)$.

\begin{figure}[ht]
    \centering
    \includegraphics[width=\columnwidth]{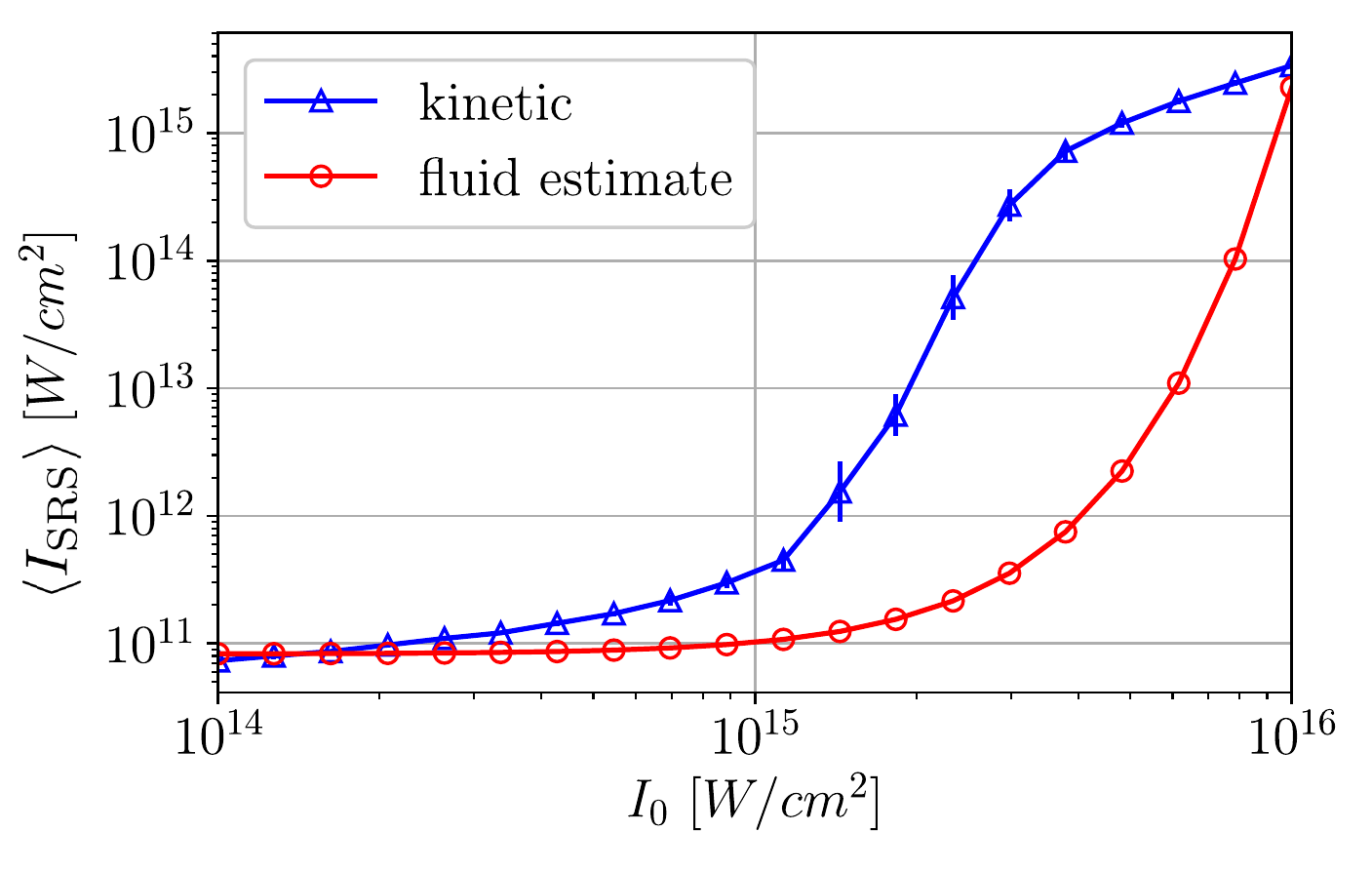}
    \caption{
        (Colour) Blue triangular markers show the intensity of SRS scattered light calculated using the fully-kinetic EPOCH code for parameters: $L_n = 500 \si{\micro\metre} $ and $n_{\mathrm{mid}} = 0.15n_\mathrm{cr}$.
        Red circular markers show the intensity of SRS scattered light calculated, for the same plasma parameters, from the fluid model presented above.
        The initial noise level in the fluid model was calculated from a PIC simulation without the laser driver: $I_\mathrm{noise}=\langle E_yB_z\rangle_{x,t} / \mu_0 = 8\times 10^{10} \si{W/\centi\metre^2}$.
    }
    \label{fig:kineticVsfluid}
\end{figure}{}

The red circular markers in Figure \ref{fig:kineticVsfluid} show the results of applying this method to the case of a $500\si{\micro\metre} $ density scale-length plasma, with the density profile centred at $0.15n_\mathrm{cr}$, for incident laser intensities ranging from $10^{14} - 10^{16}$\si{W/\centi\metre^2}. The blue triangular markers show the intensity of SRS scattered light calculated from the equivalent kinetic EPOCH simulations. The relationship between the kinetic and fluid results changes as the incident laser intensity increases. At low intensities the fluid and kinetic models are well matched, but not identical, suggesting that there is always some kinetic element to the SRS behaviour in these simulations. Continuing this analysis to intensities $<10^{13}$\si{W/\centi\metre^2} (well below those relevant to shock-ignition) shows that the two methods converge for low intensities, where the behaviour is purely fluid. Once the incident laser intensity exceeds  $I_\mathrm{threshold} \sim 1.4\times10^{15}$\si{W/\centi\metre^2}, the intensity of SRS scattered light measured in the kinetic simulations exceeds the fluid prediction by between one and three orders of magnitude, until the intensity reaches $I_0 = 10^{16}$\si{W/\centi\metre^2}, where it appears to saturate. In the fluid model, $\langle I_{\mathrm{SRS}}\rangle$ is a smooth function of incident laser intensity and we cannot define such a threshold intensity. This implies that kinetic effects in our simulations are responsible for the increase in $\langle I_{\mathrm{SRS}} \rangle$ and that we have observed iSRS. The fluid estimate shows no sign of saturating at high intensities, since the Rosenbluth gain formula used is based on unbounded linear SRS growth over the resonance region $\ell$ and the model does not include pump depletion.

By constructing plots such as these, which show the fully kinetic PIC results alongside results from our simple fluid model, we are able to identify the iSRS threshold as the point past which the kinetic and fluid models differ by at least one order of magnitude.

\begin{figure}[!ht]
 \centering
 \includegraphics[width=\columnwidth]{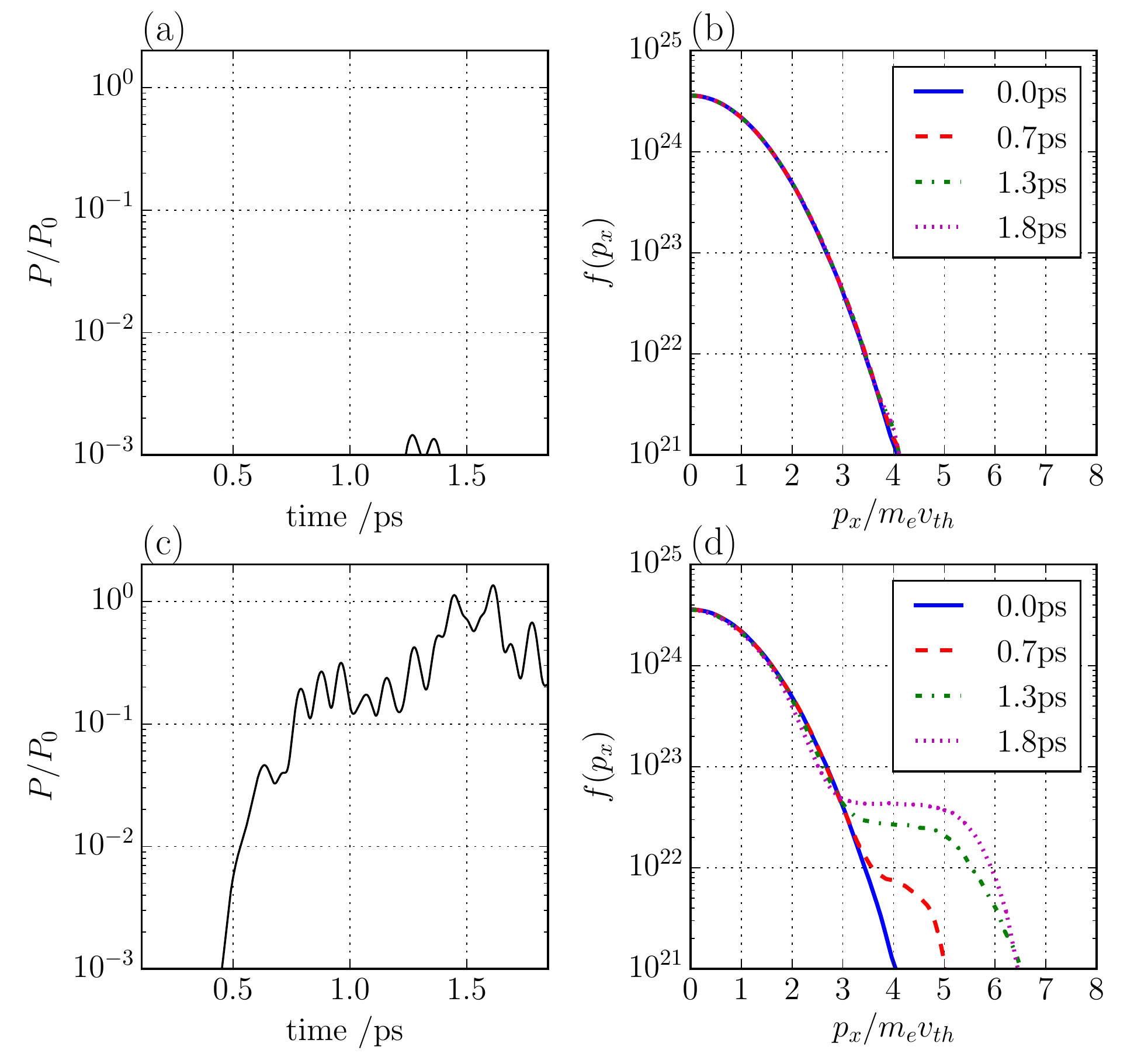}
 \caption{(Colour) Time-resolved comparison of SRS-reflectivity (a,c) and electron distribution functions (b,d) for two simulations with parameters: $L_n = 500 \si{\micro\metre} $; centred at $0.15n_\mathrm{cr}$; and $T_e = 4.5$\si{\kilo\electronvolt}. The distribution function of electron
 momentum is averaged over the entire spatial domain at four times, normalised to the initial thermal momentum. Panels (a,b) have an initial laser intensity below the threshold for inflationary SRS; $I_0 = 1.13\times10^{15}$\si{W/\centi\metre^2}. Panels (c,d) have an initial laser intensity above the iSRS threshold; $I_0 = 4.83\times10^{15}$\si{W/\centi\metre^2}.}
 \label{fig:reflAndDist}
\end{figure}

A second signature of iSRS, as reported in the literature for homogeneous plasmas, is electron-trapping in the SRS electron plasma waves, leading to a
non-linear frequency shift and enhanced SRS-reflectivies at large $k\lambda_\mathrm{D}$ \cite{Vu2002}.
A typical manifestation of this, for our inhomogeneous simulations, is shown in Figure \ref{fig:reflAndDist}. Figure \ref{fig:reflAndDist}
shows the instantaneous SRS-reflectivity measured at the left boundary of the simulation domain (a,c), alongside the box-averaged electron
distribution function at four times (b,d), for two simulations with laser intensities above and below the iSRS threshold.
Sub-figures \ref{fig:reflAndDist} (a,b) show that, when driven below threshold, the distribution of electron momenta is Maxwellian throughout the simulation, and that the maximum instantaneous power in SRS-reflected light is consequently very low ($P \sim 10^{-3}P_0$). In sub-figures (c,d), where the incident laser intensity is well above the iSRS threshold, we see that the power in SRS-reflected light is correlated with the growth of a non-Maxwellian tail in the distribution function, corresponding to an electron population trapped in the SRS electron plasma waves.
There is a general trend of increasing SRS-reflected light that correlates with
the increasing trapped electron population.

\begin{figure}[h!]
    \centering
    \includegraphics[width=\columnwidth]{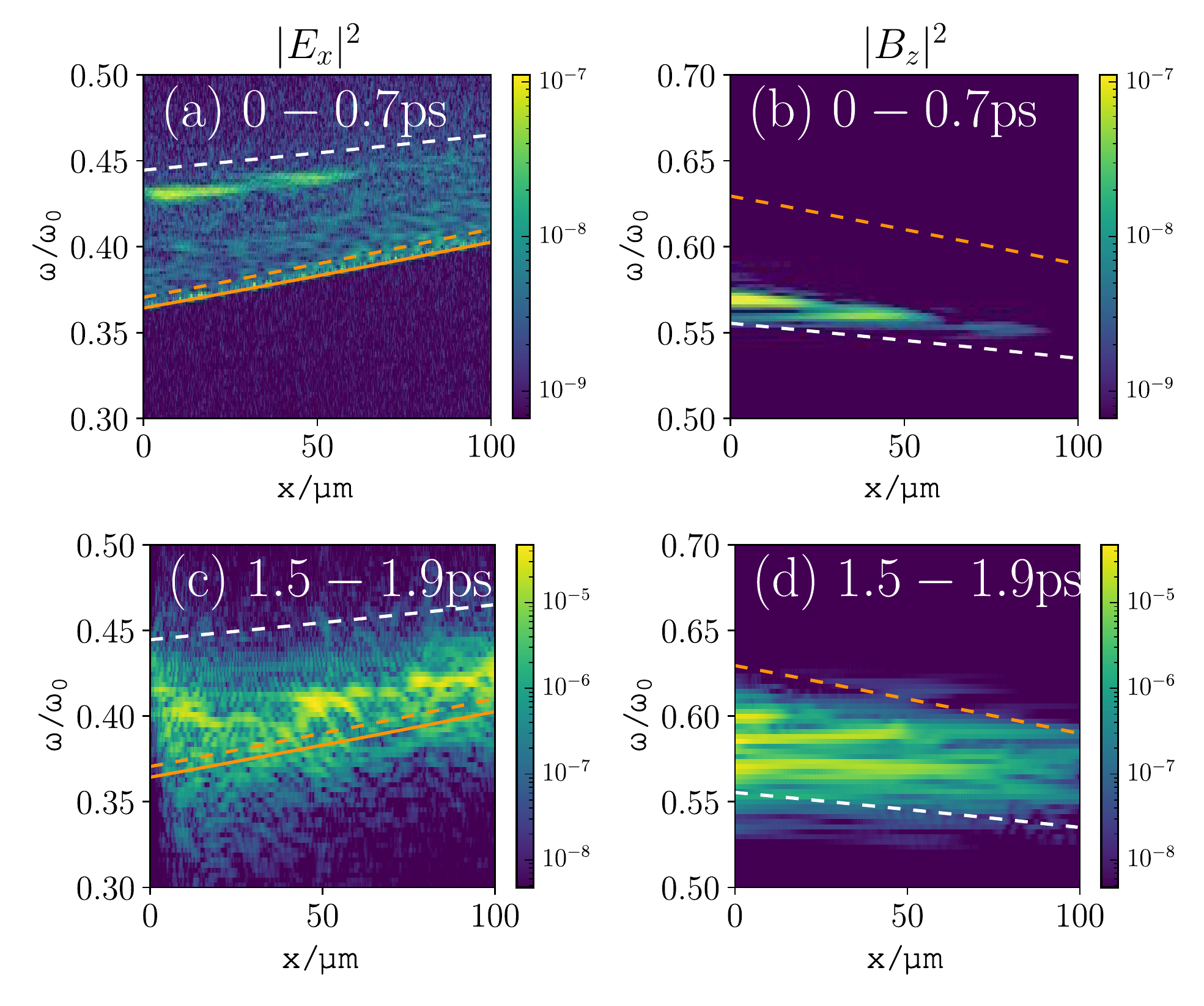}
    \caption{(Colour) Top panels show the spectra of electrostatic (a) and electromagnetic (b) waves over the period $0-0.7\si{\pico\second}$.
    The white (orange) dashed lines represent the linear predictions for the spectra of backward (forward) SRS. The bottom panels show the same spectra calculated over the period $1.5-1.9\si{\pico\second}$. The $E_x$ ($B_z$)
    spectrum is significantly down-shifted (up-shifted), demonstrating a trapped population of electrons in the EPW \cite{Yin2006}.
    The orange solid line represents the plasma frequency $\omega_{\mathrm{pe}}$ for the simulation parameters: $L_n = 500 \si{\micro\metre} $ centred at $0.15n_\mathrm{cr}$ and $I_0 = 4.83\times10^{15}$\si{W/\centi\metre^2}.}
    \label{fig:downshift}
\end{figure}

Throughout the simulation presented in Figure \ref{fig:reflAndDist}, the SRS-
reflectivity exhibits a `bursty' behaviour on the sub-picosecond timescale,
even though the distribution function appears to vary smoothly as a function of
time.
In Figure \ref{fig:reflAndDist}, the distribution functions have been averaged
over the entire simulation domain, but upon inspection of Figure
\ref{fig:downshift} (c) we can see that growth of iSRS EPWs has spatial
dependence.
Averaging the distribution over the whole domain potentially masks the
localised effects responsible for the bursty behaviour, the period of which has
been shown in Winjum {\it et al.} (2010) \cite{Winjum2010} to depend on the
local trapping-induced non-linear frequency shift.

{}

Electron-trapping in the SRS-driven EPW causes a time-dependent non-linear frequency shift of
the EPWs \cite{Morales1972,Kline2006}, and the growth of a sequence of beam-acoustic modes \cite{Yin2006}; this is the third signature of iSRS.

Figure \ref{fig:downshift} shows the spatially resolved frequency spectra of
EPWs (a,c) and EMWs (b,d) at 0-0.7ps
(a,b) and 1.5-1.9ps (c,d). In panels (a,b), the signal maxima sit very close to
the white dashed line, which represents the frequencies predicted by the SRS
matching-conditions for the original Maxwellian plasma. This means that, at
early time, the SRS EPWs and their associated back-scattered light
waves are excited at the frequencies matching those of the linear theory
without trapping.
They are slightly down-shifted from the analytical prediction, which suggests
that the trapping
becomes important almost immediately in our simulations.
At later time, Figure \ref{fig:downshift} (c) shows that the EPW spectrum is
down-shifted in frequency
at every location in the simulation domain, including to frequencies below the
plasma frequency for the original Maxwellian plasma (orange solid line).
This is evidence of a large trapped particle population removing energy from
the wave, causing the frequency of the wave
to decrease such that energy is conserved \cite{Morales1972}. We also note that
in Figure \ref{fig:downshift} (d) the back-scattered light spectrum is
up-shifted in frequency space, so as to maintain frequency matching. As well as obvious up-shift of the electromagnetic spectrum, we can also see more general broadening as we move from Figure \ref{fig:downshift} (b) to (d). This could be caused by waves from a higher density propagating to smaller $x$, so that at a particular location the spectrum covers waves from a range of densities.

\begin{figure}[ht]
    \centering
    \includegraphics[width=1.02\columnwidth]{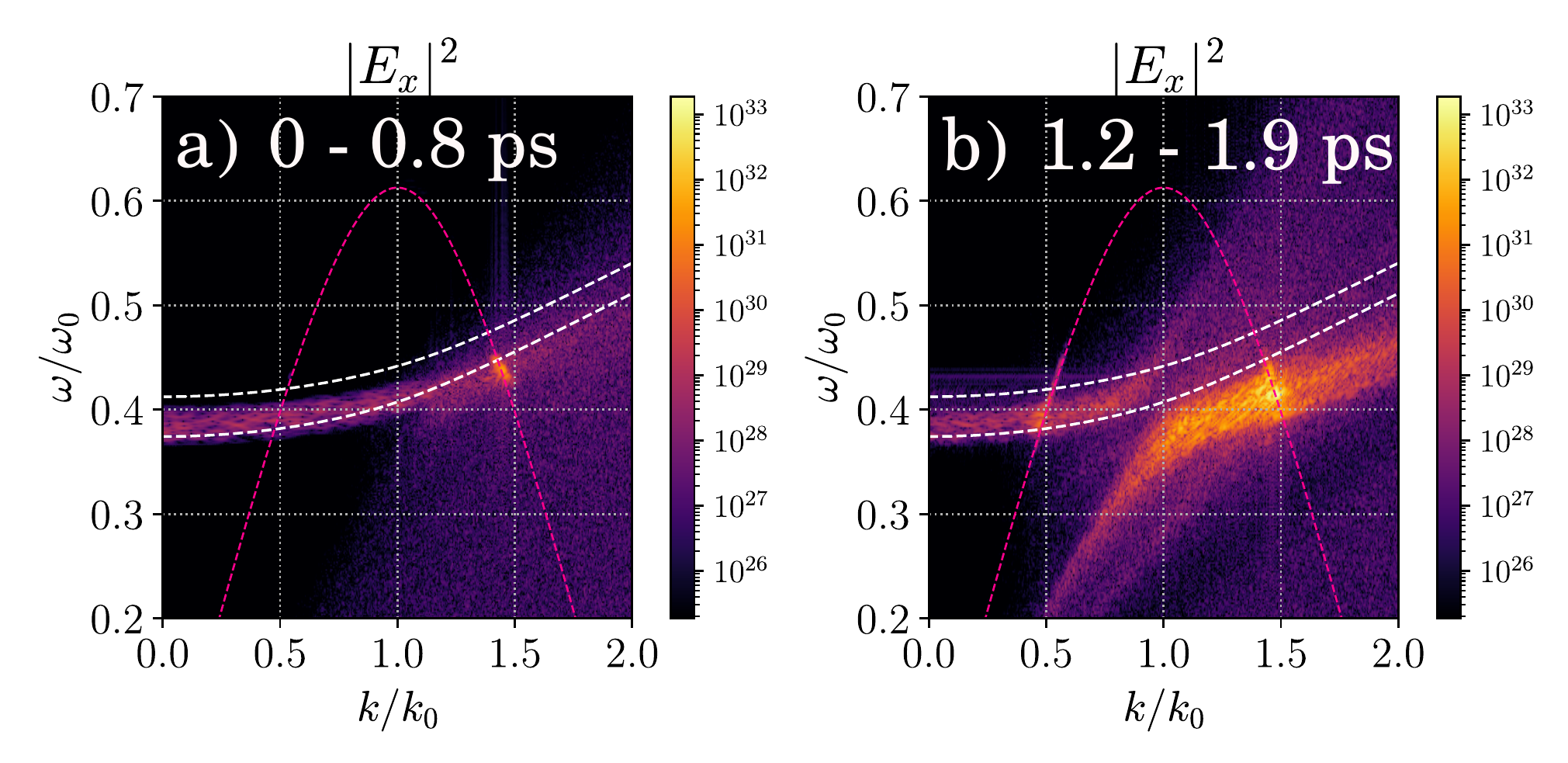}
    \caption{(Colour) (a) 2D FFT of $E_x$ over the period $ 0 - 0.8 \si{\pico\second}$. (b) 2D FFT of $E_x$ over the period $1.2 - 1.9
    \si{\pico\second}$.
    The white dashed lines represent the analytical dispersion relations corresponding to the minimum (bottom line) and maximum (top line) plasma densities,
    assuming a Maxwellian electron distribution.
    The pink dashed line shows the Stokes line for down-shifted EM waves.
    Simulation parameters: $L_n = 500 \si{\micro\metre} $ centred at $0.15n_\mathrm{cr}$
    and $I_0 = 4.83\times10^{15} \si{W/\centi\metre^2}$.
    }
    \label{fig:BAM}
\end{figure}

Further evidence for a large trapped particle population can be seen in the growth of a beam acoustic mode in the electrostatic $(\omega,k)$ spectrum. Figure \ref{fig:BAM} shows the electrostatic dispersion relation from a simulation; it is calculated by taking a 2D Fourier transform of the $E_x$ field over the entire spatial domain, and over two distinct time intervals. At early time, shown in Figure \ref{fig:BAM} (a), electron plasma waves are excited, from background noise, between the two white dashed curves. These
represent the Bohm-Gross dispersion relations $\omega_\mathrm{EPW}^2 = \omega_{\mathrm{pe}}^2 + 3v_\mathrm{th}^2k_\mathrm{EPW}^2$ for the highest density in the
domain (top line) and the lowest density (bottom line). According to fluid theory, SRS will grow where the  Stokes branch, defined by $(\omega-\omega_0)^2 = \omega_{\mathrm{pe}}^2 + c^2(k-k_0)^2$, intersects
with this dispersion curve. This fluid-SRS signal can be seen in Figure \ref{fig:BAM} (a).

{}

The right hand panel of Figure \ref{fig:BAM} shows the EPW dispersion relation calculated from the simulation between $1.2 - 1.9
\si{\pico\second}$. Inspection of the distribution function in Figure \ref{fig:reflAndDist} shows that, at these times, the distribution
function is modified from the initial Maxwellian and has a large flattened region, which acts as an effective beam population
\cite{Yin2006}. According to  linear theory, this change in the distribution function $f$ changes the kinetic dispersion relation for the electrostatic waves in the system, defined by: $\epsilon(\omega,k) = 1 - \frac{q_e^2}{\epsilon_0m_ek}\int\frac{\partial{f}/\partial{v}}{v-\omega/k} dv = 0$.
This change in the dielectric properties of the plasma is realised in the $(\omega,k)$ spectrum as a continuum of beam acoustic
modes \cite{Yin2006}, this is the large spectral feature in the right hand panel of Figure \ref{fig:BAM} that sits strictly below
the Bohm-Gross dispersion curves.
These beam acoustic modes are frequency downshifted, recovering the result from Morales and O'Neil's non-linear analysis\cite{Morales1972}. The maximum of the BAM signal at $k \sim 1.5k_0$ is the intersection of the BAM with the Stokes
branch, the new location of SRS growth.

We can also see in Figure \ref{fig:BAM} (b) a signal at $k\sim 0.5k_0$ which sits on the intersection of the Stokes branch with the range of EPWs satisfying the Bohm-Gross dispersion relations. This represents forward-scattered SRS EPWs, which have not undergone a significant frequency shift. For all the simulations presented in this paper, when driven above threshold, the power in forward SRS scattered light is of the order $P \sim 10^{-3} P_0$ or lower, and is therefore energetically unimportant.


\section{Intensity threshold and hot electron scaling}\label{sec:paramScan}
\begin{figure}[!ht]
     \centering
    \includegraphics[width=0.99\columnwidth]{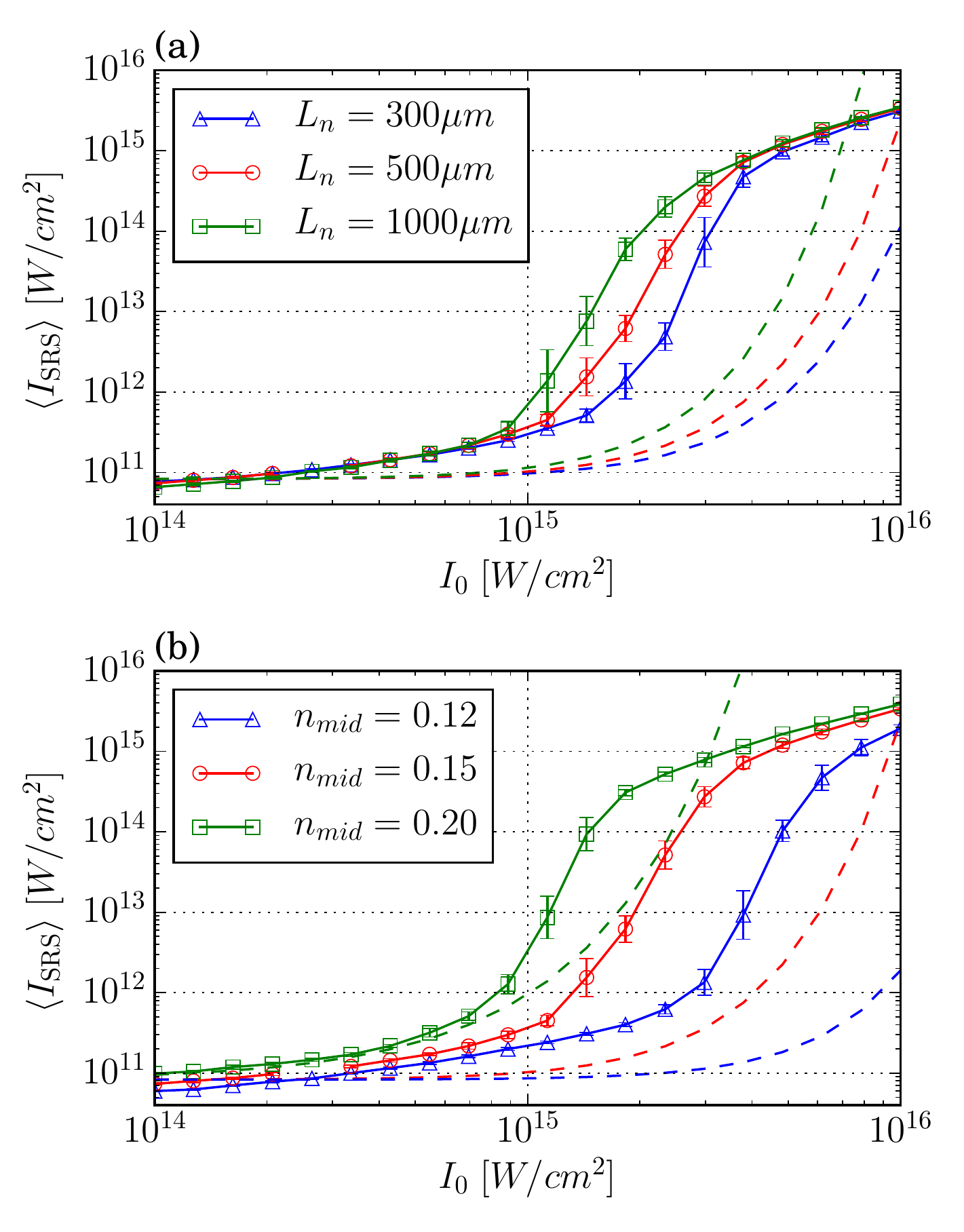}
    \caption{
    (Colour) (a)  Relationship between incident laser intensity and the intensity of SRS scattered light for three different density scale-lengths, with plasma density profiles centred at $0.15n_\mathrm{cr}$.
    (b) Relationship between incident laser intensity and the intensity of SRS scattered light for three simulations with $L_n=500\si{\micro\metre}$ centred at three different densities.
    Each coloured dashed line represents the prediction of the fluid model presented in Section \ref{sec:signatures} for the same parameters as the solid line of the same colour.
    }
    \label{fig:paramScan}
\end{figure}

Using the method developed in Section \ref{sec:signatures} for locating the inflation threshold, and the analysis of electron trapping and downshifted EPWs to ensure that the SRS observed is inflationary in origin, we investigate how iSRS depends on various plasma parameters relevant to shock-ignition. Using the PIC simulation set-up as in Section \ref{sec:code&IC} (with the same simulation domains, plasma densities, and temperatures), we varied the plasma density scale-length across the range
of values predicted for shock-ignition $(300\si{\micro\metre} - 1000\si{\micro\metre})$ \cite{Ribeyre_2009}. As well as varying the density scale-length, we also centred the density profiles at different values of density. Figure \ref{fig:paramScan} shows the result of this parameter scan.

From Figure \ref{fig:paramScan} (a) we see that as the density scale-length of the SI coronal plasma decreases, the intensity threshold for iSRS increases. Vu \textit{et al.} (2007) \cite{Vu2007} derived a condition for the kinetic inflation threshold of SRS in a homogeneous plasma. They showed that the magnitude of the trapped electron potential energy in the EPW must be greater than or equal to the energy gained by a particle in one complete trapped orbit due to velocity diffusion in the
background plasma fluctuations. This ensures that trapping remains for at least one bounce period.

No such analytic threshold has been derived for an inhomogeneous plasma.
However, when $L_n$ is smaller the inhomogenous gain is smaller and the amplitude reached by convective amplification of the SRS EPW is lower for the same intensity.
Hence SRS in a shorter density scale length plasma is less likely to generate EPWs with sufficient amplitude for electron trapping effects to trigger the transition to iSRS.

\begin{figure}[ht]
   \centering
    \includegraphics[width=\columnwidth]{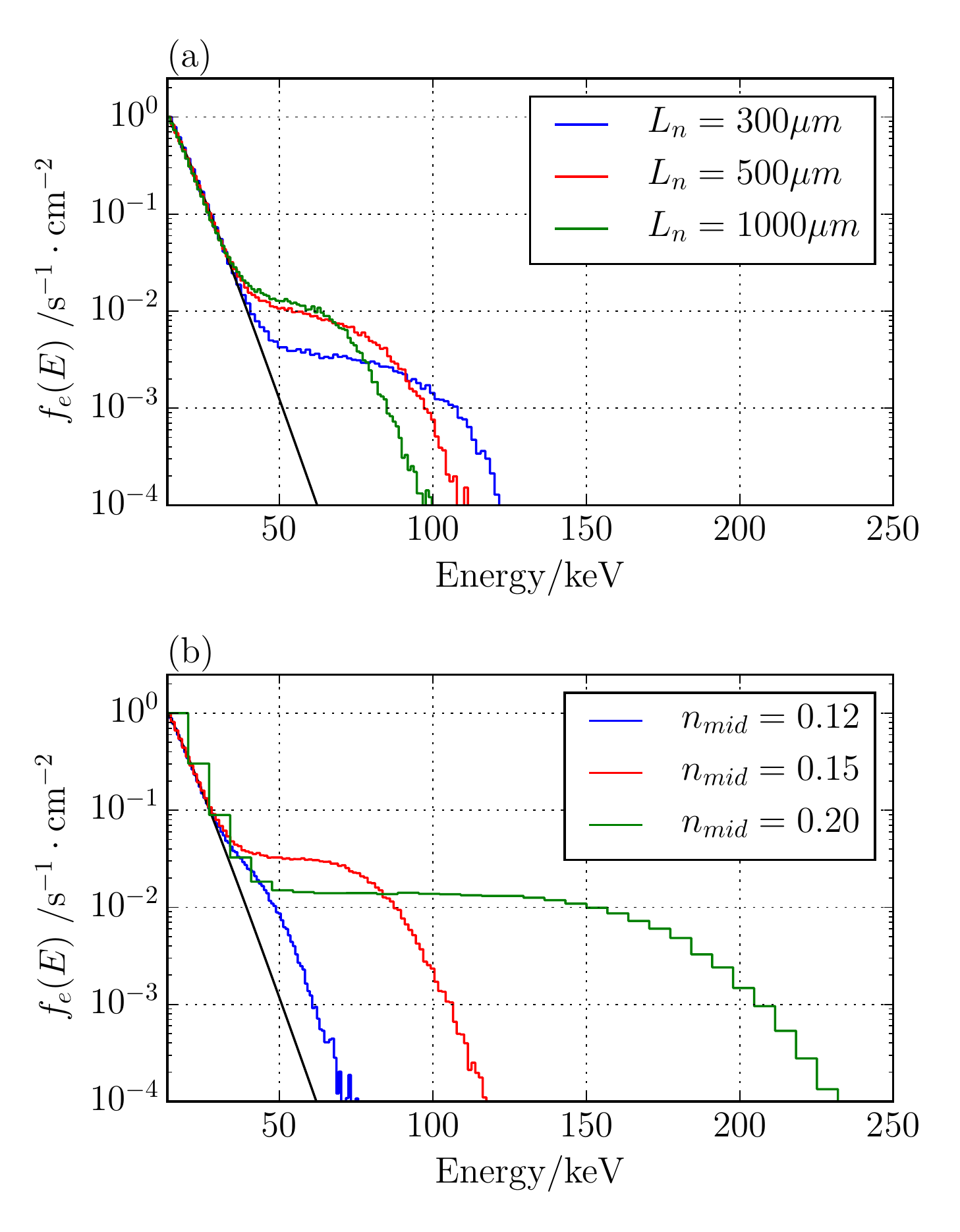}
    \caption{(Colour) (a) Hot electron flux through the right boundary in three simulations with parameters: $I_0 = 2.98\times 10^{15} \si{W / \centi \metre^2}$; $n_\mathrm{mid}=0.15 n_\mathrm{cr}$; $L_n=300,500,1000\si{\micro\metre}$.   (b) Hot electron flux through the right boundary in three simulations with parameters:
   $I_0 = 6.16\times 10^{15} \si{W / \centi \metre^2}$;  $L_n=500\si{\micro\metre}$; $n_\mathrm{mid}=0.12,0.15,0.20 n_\mathrm{cr}$ respectively. Each distribution is normalised to its maximum value. The smooth black line corresponds to the equivalent flux for a Maxwellian
   distribution with $T_e=4.5\si{\kilo \electronvolt}$, for comparison with the
   bulk plasma.}
    \label{fig:hotelectrons}
\end{figure}

Figure \ref{fig:paramScan} (b) shows the measured intensity of SRS scattered light in three sets of simulations with density profiles centred at $0.12,0.15,0.20 n_\mathrm{cr}$, chosen so that the density ranges do not overlap (see Table \ref{tab:densities}). As the central density decreases, the intensity threshold for iSRS increases. As for the case of varying scale-lengths changing the threshold, this can be explained in terms of the Rosenbluth gain\cite{Rosenbluth1972}. For a fixed density scale-length, as the density decreases, the Rosenbluth gain exponent also decreases. This means that the fluid gain through convective SRS is reduced. Hence SRS at a low density is less likely than that at higher density to generate EPWs with sufficient amplitude for electron trapping effects to trigger the transition to iSRS. For the parameters of Figure \ref{fig:hotelectrons} (b) with $I_0 = 6.16\times 10^{15} \si{W / \centi \metre^2}$, the Rosenbluth gain exponent increases from $\sim1$ to $\sim25$ as the densities increase.

{}

As well as understanding how the density scale-length of the plasma and the
density at which iSRS is driven affects the iSRS threshold, we would like to
understand how these factors affect the hot electron
population. We consider three simulations from Figure \ref{fig:paramScan} (a)
with $I_0 = 2.98\times 10^{15} \si{W / \centi \metre^2}$, and three from Figure
\ref{fig:paramScan} (b), with $I_0 = 6.16\times 10^{15} \si{W / \centi
\metre^2}$. Figures \ref{fig:hotelectrons} (a,b) show the hot electron
population in these simulations, in the form of histograms for the electron
flux through the right boundary.

Figure \ref{fig:hotelectrons} (a) shows the
electron distribution function resulting from iSRS for three density scale-lengths. These are for a laser intensity
above the onset threshold for iSRS but below an intensity which would lead to saturation. All results are
for the same central density. The most prominant difference is that the peak electron energy increases with
decreasing density scale-length. This results from the fact that the shorter density scale-length simulation
access a higher peak density since the simulation domain size is the same for all three cases. The SRS
matching conditions for these higher densities result in a higher phase speed of driven EPWs. Solving the
SRS matching conditions for these densities, we find that the hot-electron energies calculated from the phase
velocities are between $35 - 50 \si{\kilo \electronvolt}$ for all three cases.

Figure \ref{fig:hotelectrons} (b), however, shows a clear dependence of the hot
electrons from iSRS on density. As the density increases the maximum hot-
electron kinetic energy also increases in line with the increase in SRS EPW
phase velocities.
Over the 2ps of the simulations the fraction of incident laser energy converted
into hot-electrons with $\mathrm{energy} > 100\si{\kilo \electronvolt}$ are: 0,
$0.002$, and $0.15$ for the $0.12,0.15,0.20 n_\mathrm{cr}$ densities. For the
density scale-lengths $L_n=300,500,1000\si{\micro\metre}$ the fractions of
incident laser energy converted into $> 100\si{\kilo \electronvolt}$ hot-
electrons are: $0.005,0.001$ and $0$ respectively.

\section{Conclusion}\label{sec:conclusion}

Inflationary SRS has been detected in PIC simulations of a inhomogeneous plasmas
with parameters relevant to the shock-ignition model of ICF.
This study demonstrates an iSRS threshold
 $I_\mathrm{threshold} \lesssim 5\times 10^{15} \si{W/\centi\metre^2}$ across the whole range of parameters tested
 and that the location of this threshold depends on the density scale-length $L_n$. For the case with
$L_n=500 \si{\micro\metre}$ and $I_0 =4.83\times10^{15} \si{W/\centi\metre^2}$ significant iSRS  would occur
at $0.15 n_{\mathrm{cr}}$ generating hot-electrons with mostly $< 100 \si{\kilo \electronvolt}$ energies and
depleting the laser drive available at higher densities. This is potentially beneficial to
shock-ignition in that these electrons are likely to enhance the ignitor shock and prevent significant SRS at higher densities,
potentially absolute at $0.25 n_\mathrm{cr}$. SRS at higher densities is likely to
generate electron distributions with a higher percentage of
$> 100 \si{\kilo \electronvolt}$ electrons than that from iSRS at lower densities.
These results suggest that a potential route to use iSRS to the advantage of shock-ignition,
assuming all SRS cannot be removed by other means,
would be for the shock ignitor pulse to have the largest possible amplitude. This would ensure significant iSRS at lower
densities and generate only hot-electrons with energies below $100 \si{\kilo \electronvolt}$. This in turn would pump deplete the laser
reducing SRS at higher densities which could generate hot-electrons with energy above  $100 \si{\kilo \electronvolt}$. These conclusions are however only valid for the restricted 1D, collisionless simulations
presented in this paper and more detailed simulations, as outlined below, would be needed to fully assess the hot-electron
distribution and its impact on SI schemes.

The simulations presented in this paper highlight the importance of a thorough investigation of iSRS for any shock-ignition
plans. These results are however a first study of the plasmas parameters where iSRS may occur. A full theoretical
investigation of the potential impact of iSRS on shock-ignition will require significantly larger scale simulations. Of
particular importance are multi-dimensional effects and laser speckle profiles. These would allow the competition between
SRS and TPD as sources of hot-electrons to be assessed in two and three dimensions. The transverse non-uniformity associated with
multi-dimensional effects is likely to affect iSRS through trapped electron side losses and a broader spectrum of EPWs
resulting from side-scatter and TPD.
Furthermore, the auto-resonance responsible for iSRS in these simulations may not be
possible when a full speckle profile is included; since the extension of the resonance region may take the
resonant waves outside of an individual speckle. The use of broadband
laser systems to mitigate LPI will also need to be assessed
in the kinetic regime of iSRS.
All of these refinements to iSRS simulations will require considerably more
computing resources but are none-the-less needed for a comprehensive treatment of LPI relevant
to shock-ignition.

\begin{acknowledgments}
We are grateful to the EPOCH developer team (K. Bennett, C. S. Brady and H. Ratcliffe) for their adaptations
to the code in preparation for this simulation campaign.  The EPOCH code used in this work was in part funded by
the UK EPSRC grants EP/G054950/1, EP/G056803/1, EP/G055165/1, EP/ M022463/1 and EP/P02212X/1.
We also acknowledge the use of Athena at HPC Midlands+, which was funded by the EPSRC on grant EP/P020232/1,
in this research, as part of the HPC Midlands+ consortium.

This work has been carried out within the framework
of the EUROfusion Enabling Research Project: ENR- IFE19.CEA-01 ``Study of Direct Drive and Shock Ignition for
IFE: Theory, Simulations, Experiments, Diagnostics development" and has received funding from the Euratom
research and training programme. The views and opinions expressed herein do not necessarily reflect those
of the European Commission.
\end{acknowledgments}

\section*{Data availability}
Support in generating the data that support the findings of this study, using the EPOCH PIC code, are available from the corresponding author upon reasonable request.

\bibliographystyle{apsrev4-2}
\bibliography{bib}

\end{document}